\documentclass[twocolumn,aps,prd,amsmath,amssymb,preprintnumbers,longbibliography]{revtex4-1}
\usepackage{amsmath} 
\usepackage{amsfonts} 
\usepackage{amssymb}
\usepackage{bbm}
\usepackage{graphics}
\usepackage{graphicx}
\usepackage{titlesec}
\usepackage{mathtools}
\usepackage{environ}
\usepackage{dsfont}

\usepackage[colorlinks=true]{hyperref}
\hypersetup{urlcolor=black,linkcolor=black,citecolor=black}
\usepackage[toc,page]{appendix}

\usepackage{makecell,tabularx}
\setcellgapes{3pt}

\makeatletter
\renewcommand*\env@matrix[1][c]{\hskip -\arraycolsep
	\let\@ifnextchar\new@ifnextchar
	\array{*\c@MaxMatrixCols #1}}
\makeatother

\newcommand{\be}{\begin{equation}}
\newcommand{\ee}{\end{equation}}
\newcommand{\ba}{\begin{eqnarray}}
\newcommand{\ea}{\end{eqnarray}}

\newcommand{\lb}{\left}
\newcommand{\rb}{\right}
\newcommand{\tsigma}{\tilde{\sigma}}
\newcommand{\del}{\partial}

\titleformat{\subsection}[block]{\normalfont\bfseries}{\thesubsection.}{1ex}{}
\titlespacing{\section}{0pt}{20pt}{10pt}[0pt]
\titleformat*{\section}{\normalfont\bfseries}
\renewcommand{\thesubsection}{\arabic{subsection}}

\usepackage{natbib}

\usepackage{braket}

\usepackage{todonotes}

\definecolor{refkey}{rgb}{0,0,1}
\definecolor{labelkey}{rgb}{0,1,0}

\renewcommand{\thesection}{\arabic{section}}
\renewcommand{\thesubsection}{\thesection.\arabic{subsection}}

\makeatletter
\renewcommand{\p@subsection}{}
\renewcommand{\p@subsubsection}{}
\makeatother

\usepackage{subfiles}
\usepackage{enumitem}
\usepackage{mathrsfs}

\begin{document}

\title[ ]{The great emptiness at the beginning of the Universe}

\author{C. Wetterich}
\affiliation{Institut  f\"ur Theoretische Physik\\
	Universit\"at Heidelberg\\
	Philosophenweg 16, D-69120 Heidelberg}

\begin{abstract}
The great emptiness is a possible beginning of the Universe in the infinite past of physical time. For the epoch of great emptiness particles are extremely rare and effectively massless. Only expectation values of fields and average fluctuations characterize the lightlike vacuum of this empty Universe. The physical content of the early stages of standard inflationary cosmological models is the lightlike vacuum. Towards the beginning, the Universe is almost scale invariant. This is best seen by an appropriate choice of the metric field -- the primordial flat frame -- for which the beginning of a homogeneous metric is flat Minkowski space. We suggest that our observed inhomogeneous Universe can evolve from the lightlike vacuum in the infinite past, and therefore can have lasted eternally. Then no physical big bang singularity is present.
\end{abstract}

\maketitle

In most present cosmological models particles, radiation and entropy have been created in a heating period after a first ``beginning stage". This beginning stage is often described by models of cosmological inflation \cite{Starobinsky1980,Guth1981,Mukhanov1981,Linde1982,Albrecht1982,Linde1983,Shafi1983}. Other models as a bounce show similar features in the context of our discussion. In the presently dominant view the physical content of inflation is is an extremely short dramatic expansion period. In contrast, we argue here that for our observed Universe the beginning epoch corresponding to inflation extends for a very long period in physical time and may have lasted eternally \cite{Wetterich2014}. The beginning is vacuum, characterized only by average values of fields and their fluctuations. This is a very quiet epoch with only a very slow increase of particle masses. In the infinite past all particles become massless. Our present Universe with all its structures has emerged from this ``lightlike vacuum".

Alternative proposals for an eternal Universe are eternal inflation \cite{PhysRevD.27.2848, Linde:1986fd, Aguirre_2003} or the multiverse. They are based on the possibility that our observed Universe may only correspond to a small local region of a Universe with very different properties far outside our present horizon \cite{Linde1983, Shafi1983}. For eternal inflation the Universe at large is inhomogeneous, with new local inflationary Universes created continuously in many regions of the multiverse. In contrast, we discuss here the possibility that our observed Universe is part of a rather large region that can have existed with uniform properties since the infinite past. This region could be the whole Universe. We do not speculate if the Universe also could contain other regions. Those would have no impact on our observed Universe.

We base our arguments on ``field relativity" \cite{CWUWE, Wetterich2014a}, the observation that the same physical content can be described by different choices of fields. Different choices of the metric field or different ``frames" lead to different geometrical pictures. Observable quantities cannot depend on the choice of ``coordinates in field space". 
They are typically dimensionless, as the product of the distance between galaxies and the electron mass, whose change is measured by the redshift, or the ratio between the nucleon mass and the Planck mass which determines the strength of the gravitational interaction.
The observables include the properties of the primordial fluctuation spectrum, element abundances in nuclear synthesis, or the ratio between temperature and the electron mass at the emission of the cosmic microwave background. In contrast, geometric quantities as the curvature scalar or the properties of geodesics, including the issue of their completeness, are not accessible to observation. They depend on the choice of the metric frame.

We construct for standard inflationary models a ``primordial flat frame" \cite{Wetterich2014a, Wetterich2013} for which homogeneous isotropic spacetime becomes flat Minkowski space in the infinite past. This is somewhat analogous to models of ``genesis" in higher derivative theories \cite{CREM1,CREM2,RUB1,RUB2}. We remain here, however, within the standard setting of inflation models with a scalar field coupled to the metric.
Only a singular transformation of the metric to the ``Einstein frame" induces the incomplete geodesics and possible curvature singularities in the familiar picture of the big bang.
Higher derivatives are not important for the cosmological solutions discussed here.

A central outcome of our investigation is the absence of a physical big bang singularity
for the homogeneous isotropic cosmological solution. 
It has been argued that a singularity is unavoidable under rather general conditions \cite{Penrose1965,Hawking1966}, and that the lifetime of an inflationary Universe is finite \cite{Borde2001,Mithani2012}. For standard inflationary models we find that the big bang singularity of homogeneous solutions is an artifact of a singular choice of fields. This is similar (but not identical) to coordinate singularities as the south-pole singularity in Mercator coordinates. For a choice of regular field coordinates the absence of a physical singularity becomes apparent, similar to an appropriate map for Antarctica.

It is sometimes argued that a homogeneous isotropic Universe may be regular, but neighboring inhomogeneous solutions become singular. It is concluded that our observed inhomogeneous Universe has a singularity when extrapolated backwards. The present note establishes that no physical singularity needs to occur for the observed inhomogeneous Universe. If the propagators for fluctuations around Minkowski space in the primordial flat frame remain finite, our observed inhomogeneous Universe could be extrapolated backwards to the infinite past. 
We briefly discuss conditions under which the propagators of fluctuations are indeed well behaved, as well as an alternative beginning dominated by fluctuations.

\bigskip

\normalsize \textbf{Variable gravity.} \enspace
Our point can be made in models of ``variable gravity" \cite{Wetterich2013} with quantum effective action
\begin{equation}\label{1}
\Gamma = \int_x \sqrt{g} \left\{- \frac{\chi^2}{2}R + \frac{1}{2} (B-6) \del^\mu \chi \del_\mu \chi + \lambda \chi^4 \right\}.
\end{equation}
We will show by conformal field transformations or Weyl scalings \cite{HW,DIC} to the Einstein frame that such models are equivalent to standard models of inflation.
For the coefficient of the curvature scalar $R$ the fixed Planck mass $M$ is replaced by a variable Planck mass given by a scalar field $\chi$.
All particle masses are proportional to $\chi$ as well. 
For constant $\lambda$ and $B$, including vanishing values, the action contains no dimensionful parameter. Quantum scale symmetry \cite{Wetterich2019} is realized in this case. Our models will be characterized by a small violation of scale symmetry induced by a logarithmic dependence of $\lambda$ and $B$ on $\chi/\mu$ via
\begin{equation}\label{2}
x = \frac{1}{\ln \left(\frac{\mu^2}{\chi^2} + c_t\right)}.
\end{equation}
The scale $\mu$ reflects running couplings. It is the only scale in these models -- the Planck mass $M$ does not appear as an intrinsic scale. For $x \to 0$ the functions $B$ and $\lambda$ approach constants, $B(x\to 0) = 0$, $\lambda(x \to 0) = \lambda_0 > 0$. Thus quantum scale symmetry is realized for $\chi \to 0$. The infinite past will be described by a vanishing scalar field $\chi \to 0$. It is a scale-invariant Universe with massless particles. 

For $\chi \to 0$, $B \to 0$ the coefficient of the scalar kinetic term $K = B-6$ is negative. This feature is central for the existence of flat space cosmological solutions of the field equations derived from the effective action \eqref{1}. For a variable Planck mass $\chi$ the condition of stability is $B>0$. This is indeed realized and the models have no ghost -- or tachyon -- instabilities.

The quantum effective action $\Gamma$ includes all effects of quantum fluctuations. The field equations derived by variation of the quantum effective action are exact, without further quantum corrections. They determine the time evolution of expectation values in the presence of both small and large fluctuations, as bubbles or instantons. The conformal transformations underlying field relativity are simply variable transformations in a system of differential equations. This extends to propagators and the associated spectrum of primordial fluctuations~\cite{Wetterich2015}. As a price to pay, an exact computation of the quantum effective action is not possible. A given simple form as in eq.~\eqref{1} is at best a valid approximation. 

The homogeneous isotropic solutions of the field equations involve two functions of cosmic time $t$, namely the scalar field $\chi(t)$ and the scale factor $a(t)$ of a Robertson-Walker metric. 
Towards the infinite past $t\to -\infty$ the scalar field vanishes as
\begin{equation}\label{2A}
\chi(t) = \sqrt{\frac{3}{\lambda_0}}(t_0-t)^{-1}.
\end{equation}
The scale factor approaches a constant value $\bar{a}$
\begin{equation}\label{2B}
a(t) = \bar{a} \Bigg(1 + \frac{ \alpha(t)}{\ln \left(\sqrt{\frac{\lambda_0}{3}}\mu (t_0-t)\right)}\Bigg),
\end{equation} 
such that geometry becomes Minkowski space in the infinite past.
The function $\alpha(t)$ varies very slowly. Its precise form depends on the specific model for inflation and will be given below for particular models. A detailed discussion of a large family of models, field equations and their solutions can be found in an accompanying paper \cite{CWPF}.
A primordial flat frame with these propagators exists for a large class of inflationary models in the Einstein frame, both with curvature scalar divergent or finite at the bing bang singularity.
\bigskip

\textbf{Field relativity.} \enspace
Let us construct the invertible map between the action \eqref{1} in the ``scaling frame" and the standard inflationary models in the Einstein frame. In the Einstein frame the effective action describing the inflationary epoch involves the metric and a scalar ``inflaton" field $\sigma$, 
\begin{equation}\label{A3}
	\Gamma = \int_x \sqrt{g_E} \lb\{ - \frac{M^2}{2} R_E + \frac{1}{2} \partial^\mu \sigma \partial_\mu \sigma + V_E(\sigma) \rb\},
\end{equation}
with $V_E$ the effective scalar potential in the Einstein frame. A Weyl transformation,
\begin{align}\label{eq:FR2}
	g_{E,\mu \nu} = w^2 g_{\mu \nu}, \quad w^2 = \frac{\chi^2}{M^2},
\end{align}
relates the metric $g_{E, \mu \nu}$ in the Einstein frame and the metric $g_{\mu\nu}$ in the scaling frame. The scalar field $\chi$ will be related to $\sigma$.

Expressed in terms of $g_{\mu \nu}$ one obtains the action \eqref{1} of ``variable gravity"\cite{Wetterich2013} with
\begin{align}\label{eq:FR3}
	\lambda(\chi) = \frac{V_E(\tilde{\sigma})}{M^4}, \quad B(\chi) = \chi^2 \lb( \frac{\partial \tilde{\sigma}}{\partial \chi} \rb)^2, \quad \tilde{\sigma} = 
	\sigma / M.
\end{align}
We assume a monotonic behavior $B>0$. During inflation the $\chi$-dependence of $\lambda$ is directly related to the slow roll parameter $\epsilon$,
\begin{equation}\label{eq:FR6}
\lb( \frac{\partial \ln\lambda}{\partial \ln\chi}\rb)^2 = 2B\epsilon, \quad \epsilon = \frac{1}{2} \lb( \frac{\partial \ln V_E}{\partial \tsigma}\rb)^2 .
\end{equation}
Starting with the action \eqref{1} for variable gravity the Planck mass $M$ appears only through the definition of the fields $g_{E,\mu\nu}$ and $\sigma$. It is not an intrinsic scale of the theory. Field relativity states that all expectation values of observables computed from the actions \eqref{1} and \eqref{A3} are the same.  

\bigskip

\textbf{Primordial flat frame.} \enspace
At this stage we still have a whole family of frames according to different possible choices for the relation between $\sigma$ and $\chi$. Many models admit a ``primordial flat frame" by a choice of $\tsigma(\chi)$ for which the ``kinetial" $K = B-6$ obeys
\begin{align}\label{eq:FR7}
	K<0, && K+6 = \frac{\partial \ln K}{\partial \ln \chi} - \frac{\partial \ln \lambda}{\partial \ln \chi}.
\end{align}
With the choice \eqref{eq:FR7} there are cosmological solutions for which spacetime is flat, while the dynamics of inflation is associated to the evolution of $\chi$.

The metric field equations derived from the action \eqref{1} read for a Robertson-Walter metric $(R = 12 H^2 + 6\del_t H$, $H=\del_t \ln(a))$
\begin{equation}\label{eq:FR9}
	3 \chi^2 H^2 = \lambda \chi^4 + \frac{K}{2} \dot \chi^2 - 6H \chi \dot \chi,
\end{equation}
and
\begin{equation}\label{eq:FR10}
\chi^2 R = 4 \lambda \chi^4 - (K+6)\dot \chi^2 - 6 \chi ( \ddot \chi  + 3 H \dot \chi).
\end{equation}
Here dots denote time-derivatives.
The scalar field equation is given by
\begin{equation}\label{eq:FR8}
K(\ddot \chi + 3H \dot \chi) = - 4 \lambda \chi^3 - \chi^4 \frac{\del \lambda}{\del \chi} + \chi R - \frac{1}{2} \frac{\del K }{\del \chi} \dot \chi^2.
\end{equation}

For frames obeying the condition \eqref{eq:FR7} all three equations \eqref{eq:FR9}--\eqref{eq:FR8} can be solved for a flat Minkowski geometry. For this solution one has
\begin{align}\label{eq:FR11}
\dot \chi = c(\chi)\chi^2, &&  c= \sqrt{ -\frac{2\lambda}{K}}, && H=0, && R=0.
\end{align}
If $c$ reaches a 
constant for $\chi \to 0$, one finds for the asymptotic behavior in the past infinity $t\to -\infty$ that $\chi$ vanishes according to
\begin{align}\label{eq:FR13}
\chi(t) \to \frac{1}{c(t_0-t)+\chi_0^{-1}}.
\end{align}
We typically encounter slowly varying functions $c(\chi)$ for which eq.~\eqref{eq:FR13} remains a good approximation.

The solution in the primordial flat frame has a regular geometry. This demonstrates the absence of a physical singularity 
for this cosmological solution.
The singularity in the Einstein frame is a field singularity induced by the singularity in the field transformation \eqref{eq:FR2} for $\chi \to 0$.
While the metric $g_{\mu\nu}$ amounts to ``regular field coordinates" for the infinite past, the Einstein metric $g_{E,\mu\nu}$ corresponds to ``singular field coordinates". The regular field coordinates provide for a more natural description of the physical properties of the lightlike vacuum. Conformal time is the same for both frames.  As $\chi(t \to -\infty) \to 0$ all particles become massless in the infinite past. 

We next establish the existence of a frame for which the condition \eqref{eq:FR7} holds. This ``flat frame condition"~\eqref{eq:FR7} constitutes a differential equation for the function $B(\tsigma)$  
\begin{equation}\label{eq:FR14}
B=2\epsilon \lb( 1 \pm \frac{1}{\sqrt{2\epsilon}(6-B)} \frac{\del B}{\del \tsigma} \rb)^2.
\end{equation}
In eq.~\eqref{eq:FR14} the plus sign applies if $V_E$ increases with $\sigma$, while the minus sign accounts for $V_E$ decreasing with $\sigma$. A given solution $B(\tsigma)$ of eq.~\eqref{eq:FR14} determines the relation between $\tsigma$ and $\chi$ by eq.~\eqref{eq:FR3}, resulting in $B(\chi)$ and $\lambda(\chi)$. A primordial flat frame exists whenever for a given $V_E(\tsigma)$ and associated $\epsilon(\tsigma)$ a solution of eq.~\eqref{eq:FR14} with $0<B(\tsigma)<6$ exists. In particular, for constant $\epsilon$ one has constant $B= 2\epsilon \ll 1$, such that $K=B-6$ is indeed negative. For small $\epsilon$ one can solve eq.~\eqref{eq:FR14} iteratively
\begin{equation}
B = 2\epsilon \lb( 1 \pm \frac{1}{3-\epsilon} \sqrt{\frac{\epsilon}{2}} \frac{\del \ln{\epsilon}}{\del \tsigma} \rb)^2.
\end{equation}
We will only require that $\chi(\tsigma)$ is defined such that the condition \eqref{eq:FR7} holds with sufficient accuracy for $\chi \to 0$. In this case one finds solutions that approach flat space in the infinite past and are again free of singularities.

If the condition~\eqref{eq:FR7} holds the field equations admit a particular homogeneous isotropic solution that ``begins" with Minkowski space. Geodesics are complete and both conformal and cosmic time (proper time of comoving observer) can extend to minus infinity, somewhat similar to particular cosmologies with non-zero spatial curvature~\cite{Ellis:2002we}.
Our homogeneous solutions with a flat- space beginning are cosmic attractor solutions. One may envisage a beginning in the far distant past with different initial conditions, as inhomogeneous space-times or homogeneous space-time with non-zero spatial curvature. The corresponding solutions approach the attractor solution as time increases. The only instability is the slowly increasing scalar field. For many different possible beginnings the physical properties of the major part of very early cosmology are well described by ``great emptiness". This includes the epoch relevant for the characteristics of the primordial fluctuation spectrum.

\bigskip

\textbf{Starobinski inflation.} \enspace
As an example we discuss Starobinski inflation \cite{Starobinsky1980}. At present this model is compatible with all observations. By use of suitable variables~\cite{Whitt:1984pd,CWME} Starobinski inflation is characterized in the Einstein frame \eqref{A3} by a potential
\begin{equation}
V_E = \frac{M^4}{8C} \left[1 - \exp \left(-\sqrt{ \frac{2}{3}}\tsigma\right)\right]^2.
\end{equation}
For the relation between $\tsigma = \sigma/M$ and $\chi$ we employ
\begin{equation}
W = \exp \left(-\sqrt{ \frac{2}{3}}\tsigma \right) = \frac{3x}{2} \left(1-\frac{5}{6} x \ln \left( \frac{2}{3x}\right)\right),
\end{equation}
with $x$ given by eq.~\eqref{2}. For the effective action \eqref{1} this implies
\begin{equation}
\lambda = \lambda_0 (1-W)^2, \quad \lambda_0 = \frac{1}{8C},
\end{equation}
and
\begin{equation}
B = 6 x^2 \left[1 - \frac{5x}{3} \left(\ln \left(\frac{2}{3x}\right) -1\right)\right].
\end{equation}
One may verify that the primordial flat frame condition \eqref{eq:FR14} is obeyed up to terms of the order $W^4$.

The functions $B(x)$ and $\lambda(x)$ specify the action \eqref{1} for variable gravity.
Solving the field equations \eqref{eq:FR9}-\eqref{eq:FR8} for this model with $K=B-6$, one finds indeed eqs.~\eqref{2A},\eqref{2B}, with $\alpha$ approximated by
\begin{equation}
\alpha(t) = \frac{41}{48} \ln \left[ \frac{2}{3}\ln \left(\frac{\mu^2 \lambda_0 (t_0-t)^2}{3}\right) \right].
\end{equation}
Details can be found in an accompanying paper \cite{CWPF}, where we also show that the same $\chi(\eta)$ solves the field equations in the Einstein frame and in the primordial flat frame. In this paper we further discuss the primordial flat frame for other inflationary models.

\bigskip

\normalsize \textbf{Lightlike vacuum.} \enspace
Physical properties do not depend on the choice of frames. For a discussion of observables we therefore concentrate on dimensionless quantities that are invariant under Weyl scalings. As a first physical property one finds that all particles become massless as one approaches the infinite past. Massless particles indicate unbroken scale symmetry. The relevant dimensionless quantity is the ratio of particle mass $m$ over momentum $p$. For $m/p \to 0$ particles are relativistic and propagate like light.

In the primordial flat frame the lightlike behavior of all particles is seen directly. The physical momentum $p$ is given in terms of the comoving momentum $k$ as $p = k/a$. With $a$ approaching a constant $\bar{a}$, both quantities are proportional to each other. For $t\to -\infty$ and finite $k$, $p$ also remains finite. On the other hand, quantum scale symmetry implies that all particle masses are precisely proportional to $\chi$. They vanish for $t\to -\infty$ since $\chi\to 0$. 

Particle masses $\sim \chi$ in the scaling frame correspond to fixed particle masses $m$ in the Einstein frame. In the Einstein frame the momentum diverges for any $k\neq 0$ as the singularity is approached for $a_E\to 0$ .
Again $m/p$ goes to zero and all particles become effectively massless. 
On a technical level this property is directly seen in the standard analysis of the evolution of massive particle fluctuations in conformal time, see below.
The time of the big bang singularity in the Einstein frame corresponds to the infinite past in the scaling frame.

\bigskip

\normalsize \textbf{Physical time.} \enspace
We have to define a notion of ``physical time" that is the same in both frames. Cosmic time $t$ in the Robertson-Walker metric depends on the choice of frame. 
The same holds for proper time \cite{Wetterich2014}. Geometry is geodesically incomplete in the Einstein frame, while it is geodesically complete in the primordial flat frame.
Once a physical time is established, one can map physical time to cosmic time or proper time in each frame separately. Proper time cannot be employed for massless particles. Since particles become massless in the infinite past or at the big bang singularity, proper time does not seem suitable for a physical time for this period -- for details see ref.~\cite{Wetterich2014}.

Physical time should be based on oscillatory phenomena and a counting of oscillations. It is no accident that some type of ``oscillation time" has been employed since the earliest descriptions of nature by humans. Today we use it by counting the oscillations of photons with a wavelength given by some particular atomic transition. The number of oscillations  of the photon wave function with a given comoving momentum $k$ remains a valid physical time for all epochs of the Universe, including the beginning. Since the counting is discrete, it does not depend on the choice of coordinates. Neither does it depend on the choice of fields or the metric frame. 

Expressed in terms of conformal time $\eta$, the wave equation for a massless particle in a homogeneous isotropic Universe reads 
\begin{align}\label{PT1}
\left( \partial_\eta^2 + k^2 - \frac{a^2R}{6}\right) a \varphi_k = 0,
\end{align}
with $\varphi_k$ an appropriate component of the wave function in an eigenstate of comoving momentum $k$, $H = \partial_t \ln a,$ $a d\eta = dt$, and $R$ the curvature scalar. For $|a^2 R| \ll k^2$, which holds at the beginning of inflation, the number of oscillations $n_k$ is proportional to conformal time,
$n_k = \frac{k \eta}{2\pi}.$
We can therefore consider conformal time $\eta$ as a good proxy for oscillation time. For homogeneous isotropic cosmologies we can take it as physical time. Conformal time is invariant under conformal transformations of the metric and therefore the same in all frames related by Weyl scaling.
For a massive particle in the Einstein frame one adds in the bracket in eq. \eqref{PT1} a term $a^{2}m^{2}$. It vanishes for $a\rightarrow 0$.

In the scaling frame with $a \to \bar{a}$ cosmic time $t$, conformal time $\eta$ and oscillation time $n_k$ are all proportional to each other. Physical time can be extended to the infinite past if the proposed cosmological solution describes the Universe for $t \to -\infty$. In the Einstein frame, physical and conformal time are the same as in the scaling frame.
For typical inflationary cosmologies without a ``beginning event" both conformal time $\eta$ and oscillation time $n_k$ go to minus infinity as $a_E \to 0$. The Universe exists therefore since the infinite past if physical time is used -- it is eternal. Only the mapping to proper time becomes singular for $a_E \to 0$, as may be expected for particles becoming massless. (See ref. \cite{Wetterich2014} for a discussion of physical time fore massive particles.) Measured in proper time the duration of oscillations approaches zero very rapidly for $a_E \to 0$, whereas the number of oscillations goes to infinity. While the proper time interval between two ticks of the ``photon clock" is frame dependent, the number of ticks is not. 
\bigskip

\normalsize \textbf{Inhomogeneous Universe.} \enspace
Our Universe is not homogeneous and isotropic. The question arises if our observed inhomogeneous Universe can have lasted since ever in physical time, or if the extrapolation backwards necessarily encounters a physical singularity. It is often believed that the latter is the case and therefore a physical big bang singularity is unavoidable in presence of the observed inhomogeneities. We will argue here that the observed inhomogeneities may be compatible with a Universe existing since infinite physical time, with a big bang singularity being a field singularity similar to the homogeneous and isotropic solution. 
For this purpose we connect the possible divergencies of neighboring inhomogeneous cosmological solutions to the form of the propagator for the corresponding particles. For a well behaved propagator for the graviton and scalar fluctuations our inhomogeneous Universe can indeed evolve from an inhomogeneous fluctuating lightlike vacuum in the infinite past, with average inhomogeneities given by the fluctuations of the corresponding fields. A well behaved graviton propagator for $\chi\rightarrow 0$ may require, however, terms in the effective action beyond the ansatz \eqref{1}, typically involving higher derivatives of the metric.

We expand the metric around a homogeneous isotropic averaged metric, 
\begin{equation}\label{eq:IU1}
g_{\mu \nu}(\eta,x) = a^2(\eta)(\eta_{\mu\nu}+ \gamma_{\mu\nu}(\eta,x)),
\end{equation}
with $x \in \mathbb{R}^3$ denoting comoving spacelike coordinates, and similar for the scalar field
$\chi(\eta,x) = \bar{\chi}(\eta) ( 1+\delta(\eta,x)).$
The Weyl scaling \eqref{eq:FR2} relates the scaling frame to the Einstein frame
\begin{equation}\label{eq:IU3}
a^2(\eta_{\mu\nu} + \gamma_{\mu\nu}) = \frac{M^2 a_E^2}{\bar{\chi}^2(1+\delta)^2}(\eta_{\mu\nu} + \gamma_{E\mu\nu}).
\end{equation}
With  $a(\eta) = (M/\bar{\chi}(\eta)) a_E(\eta)$ one has in linear order
$\gamma_{\mu\nu} = \gamma_{E,\mu\nu} - 2 \delta \eta_{\mu\nu}.$
We concentrate here on the graviton or traceless transverse tensor fluctuations $\gamma_{mn}$, $m,n=1...3$. They obey 
$\gamma_{mn} \delta^{mn} =0,$ $k^m \gamma_{mn} = 0,$
where we have switched to a Fourier representation $\gamma_{mn}(\eta,k)$, with $k_m$ the spacelike comoving momentum, $k^m = \delta^{mn}k_n$, $k^2 = k^m k_m$. The relative graviton fluctuations $\gamma_{mn}$ are invariant under conformal frame transformations, $\gamma_{mn} = \gamma_{mn,E}$.

For the effective action \eqref{1} the linearized field equations for $\gamma_{mn}(k)$ can be written in a frame invariant form \cite{Wetterich2015} 
\begin{equation}\label{eq:IU6}
(\del_\eta^2 + 2 \hat{\mathscr{H}} \partial_\eta + k^2) \gamma_{mn}(\eta,k) =0, 
\end{equation}
with
$\hat{\mathscr{H}} = \mathscr{H} +\frac{1}{2} \del_\eta \ln F$,  $\mathscr{H}=\del_\eta \ln a$. Here $F(\chi)$ is the coefficient multiplying the curvature scalar in the effective action. In the scaling frame one has $F = \bar{\chi}^2$, while in the Einstein frame $F = M^2$.
The frame invariant formulation \eqref{eq:IU6} allows us to take over the solution for $\gamma_{mn}(\eta,k)$ from the scaling frame to the Einstein frame and vice versa. In particular, if $\gamma_{mn}(\eta \to - \infty)$ and its derivatives remain finite, there is no physical singularity for this type of inhomogeneous cosmologies. 

The general solution of eq.~\eqref{eq:IU6} reads
\begin{equation}\label{IN2}
\gamma_{mn}(k) = c_{mn}^-(k) w_k^-(\eta) + c_{mn}^+(k)w_k^+(\eta).
\end{equation}
The mode functions infered from the effective action \eqref{1} are given by
\begin{equation}\label{IN3}
w_k^-(\eta) = (w_k^+(\eta))^* = \frac{1}{A\sqrt{2k}} \left(1-\frac{i}{u}\right)e^{-iu}.
\end{equation}
They involve the frame invariant quantities
\begin{equation}\label{IN4}
A = a\sqrt{F}, \quad u=k(\eta - \eta_0),
\end{equation}
such that the solution \eqref{IN2}, \eqref{IN3} is valid in arbitrary frames \cite{Wetterich2015}. In the primordial flat frame one has $A^{-1} \approx \tilde{c}(\eta_0 - \eta) = - \tilde{c} u/k$, while for the Einstein frame $A^{-1} \approx H_E(\eta_0-\eta)/M$. The function $\tilde{c} = \dot \chi/\chi^2$ generalizes $c$ in eq.~\eqref{eq:FR11} -- it equals $c$ for $\eta \to -\infty$. The particular mode functions \eqref{IN3} obtain for $\del_\eta \hat{\mathscr{H}}/\hat{\mathscr{H}}^2 = 1+\nu$ and $\nu = 0$. Their generalization for constant $\nu \neq 0$ can be found in ref.~\cite{Wetterich2015}. Within the validity of the linear approximation the inhomogeneous graviton-like cosmological solutions \eqref{IN2} are damped oscillations. Their amplitude is frozen for $|u|\ll 1$. Early relative inhomogeneities in the graviton sector tend to become smaller as time increases. The homogeneous solution is an attractor in this sense. 

Towards the beginning the combination $A$ approaches zero both in the primordial flat frame $(F\to 0)$ and the Einstein frame $(a_{E}\to 0)$ . Eqs~\eqref{IN2},\eqref{IN3} imply that the relative fluctuations diverge for $\eta\to -\infty$ . This implies a breakdown of the linear approximation. The Universe starts in a state for which inhomogeneous configurations dominate. A frame-invariant metric can be defined as $\tilde{g}_{\mu\nu}=Fg_{\mu\nu}$ . In the linear approximation the frame invariant metric for the graviton vanishes $\sim A$, in contrast to $A^{2}$ for the homogeneous metric. This may suggest an inhomogeneous beginning for which the expectation value of the frame-invariant metric vanishes. There may exist better adapted definitions of a metric in this case.

\bigskip

\normalsize \textbf{Graviton propagator.} \enspace
The mode functions are directly connected to the propagator $G_{grav}$ for the relative graviton fluctuations, and in turn to the observable primordial tensor spectrum $\Delta_{T}^2$,
\begin{equation}\label{IN5}
G_{grav}(k,\eta) = 4|w_k^-(\eta)|^2, \quad \Delta_T^2(k) = \frac{k^3 G_{grav}(k,\eta)}{\pi^2}.
\end{equation}
As long as the graviton propagator remains finite, the mode functions remain finite and the inhomogeneous solutions can be extrapolated towards the infinite past without encountering any singularity. The same holds for the gauge invariant scalar fluctuations. As long as the scalar propagator and the associated scalar primordial fluctuation spectrum remain finite no singularity can occur in this sector.

The graviton propagator is the inverse of the second functional derivative of the effective action with respect to the graviton fluctuations. The graviton propagator is finite for $k \neq 0$ for a very extended epoch when $\chi > 0$, independently of the details of the beginning. For this epoch the description of graviton fluctuations in almost flat space becomes very simple in the primordial flat frame.

For the effective action \eqref{1} in the primordial flat frame the inverse graviton propagator is proportional to $\chi^2$. As a consequence, the graviton propagator diverges for $\chi \to 0$, as visible in eqs.~\eqref{IN3},\eqref{IN5}. 
Correspondingly, also the neighboring inhomogeneous cosmologies could become singular for the field equations derived from the action \eqref{1}. This could be related to the singular inhomogeneous cosmologies found in earlier studies on the Einstein frame \cite{LIKA, STA2}. We recall, however, that the divergence of the propagator concerns the relative metric fluctuations, while the fate of the fluctuations in the frame-invariant metric could be different. 

The graviton propagator is a direct measure for the fate of small deviations from a homogeneous Universe in the corresponding sector. For the effective action \eqref{1} the damping of the relative fluctuations according to the mode functions \eqref{IN3} is so strong that infinitely strong relative fluctuations are needed at initial time $\eta\rightarrow -\infty$ in order to produce the predicted primordial graviton fluctuations during later stages of inflation. Finite relative inhomogeneities would be completely wiped out before the end of inflation. No such issue occurs if initial conditions are set at some finite initial time.

Even if the relative fluctuations diverge, this is not per se a problem. Finite inhomogeneities in the frame-invariant metric $\tilde{g}_{\mu\nu}$, paired with a vanishing expectation value for the homogeneous solution, necessarily lead to diverging relative inhomogeneities.

The divergence of the (relative) graviton propagator could be an artifact of an insufficient approximation to the quantum effective action. One expects the presence of higher derivative terms \cite{Wetterich2019}
\begin{equation}\label{IN6}
\Gamma_{hd} = \frac{1}{2}\int_x \sqrt{g} \left\{ C^{\mu\nu\rho \sigma} D C_{\mu\nu\rho \sigma} -RCR \right\},
\end{equation}
with Weyl tensor $C_ {\mu\nu\rho\sigma}$. Here $D$ and $C$ are dimensionless functions of the covariant Laplacian divided by some squared renormalization scale $k^{2}$, as well as of $\chi^{2}/k^{2}$. (There are other possible terms as well.)
For $\chi = 0$ these terms typically dominate the inverse propagator. They can render the graviton propagator finite for $k \neq 0$. 
(See ref. \cite{wetterich2020fundamental} for an example.) For a graviton propagator $G_{grav}(k,\eta)$ that remains finite for all $\eta$  the mode function $w_{k}^{-}(\eta)$ in eq. \eqref{IN5} no longer diverges for $\chi\rightarrow 0$. Correspondingly, the relative graviton perturbation $\gamma_{mn}$ in eq. \eqref{IN2} remains finite if we insert the coefficient $c_{mn}^{-}(k)$ that corresponds to the primordial tensor fluctuations in our Universe.
If the inhomogeneous solutions dominating the primordial fluctuation spectrum remain small deviations from the homogeneous ``background" solution, these neighboring inhomogeneous cosmologies will remain finite for all $\eta$, even in the infinite past.

For the alternative of diverging relative fluctuations, the beginning of the Universe could be a strongly fluctuating state with $\bar{\chi}=0$. The propagator for the relative graviton fluctuations in a homogeneous background is no longer the relevant physical quantity in this case. Approximate homogeneity would be reached only once $\bar{\chi}$ has grown sufficiently large.

\bigskip

\normalsize \textbf{Higher derivative invariants.} \enspace
Higher derivative terms of the type \eqref{IN6} are relevant for understanding the graviton propagator for $\chi = 0$. In the primordial flat frame they play only a minor role for the homogeneous cosmological solutions. The Weyl tensor vanishes for the homogeneous solutions and 
the first term in eq. \eqref{IN6} 
does not contribute to the homogeneous field equations. The ratio $R/\chi^2$ vanishes for $\eta \to -\infty$, such that the relative importance of the term $\sim R^2$ goes to zero in this limit.

General inhomogeneous solutions typically lead to a non-zero Weyl tensor. In the Einstein frame it is often observed that the squared Weyl tensor diverges for $a_E \to 0$, and this is incorrectly associated with a physical singularity. In the primordial flat frame a non-zero finite Weyl tensor for $\eta \to -\infty$ implies that the combination $W=\sqrt{g}C^{\mu\nu\rho\sigma}C_{\mu\nu\rho\sigma}$ differs from zero. 
The quantity $W$, which includes the factor $\sqrt{g}$, is invariant under Weyl scalings. It is therefore the same in the Einstein frame. This implies that in the Einstein frame the squared Weyl tensor indeed diverges $\sim 1/\sqrt{g_E}$ as $\sqrt{g_E}$ reaches zero at the ``big bang singularity". (For the homogeneous solution and conformal time one has $\sqrt{g} = a^4 \to \bar{a}^4$, $\sqrt{g_E} = a_E^4 \to 0$.) The regular behavior in the primordial flat frame demonstrates that the apparent singularity is a field singularity, arising from a singular choice for the metric field. No physical singularity 
can be infered from the diverging squared Weyl tensor alone, in contrast to the combination $W$.
\bigskip

\normalsize \textbf{Prediction for decreasing modes.} \enspace
Frame-invariant fields can be divided into observable and decreasing modes. Decreasing modes $\Delta$ are eigenvectors of the stability matrix for a linearized evolution equation
\begin{equation}
\del_\eta \Delta = \zeta \Delta, \quad \zeta < 0.
\end{equation}
Here $\Delta$ is typically a small deviation from some family of attractor solutions.
The solution for constant $\zeta$,
\begin{equation}
\Delta(\eta) = \Delta(\eta_{in})\exp\{\zeta(\eta - \eta_{in})\},
\end{equation}
decreases for increasing $\eta - \eta_{in}$ $(\Delta(\eta_{in}) > 0)$. For a bounded initial value $\Delta(\eta_{in})$ and initial time $\eta_{in}\to -\infty$ one finds $\Delta(\eta) \to 0$. A Universe lasting since ever makes the prediction that all decreasing modes are zero. The general analysis of small cosmic fluctuations indicates that indeed some of the modes are decreasing modes in this sense. On the other hand, the modes dominating the primordial fluctuation spectrum cannot be decreasing modes in this sense. The frame-invariant graviton field increases $\sim A$ and therefore corresponds to an observable mode.

The presence of decreasing modes implies that arbitrary field configurations at some finite $\eta$ cannot be extrapolated backwards to arbitrarily large negative $\eta$ without encountering a singularity. This includes metric configurations in the close vicinity of the observed metric of the Universe. The reason is the prediction for the allowed range of field values at finite $\eta$ if initial conditions are set in the infinite past. Field values outside the predicted range are inconsistent and lead to a singular behavior if one tries to extrapolate them backwards.

If one extrapolates backwards non-zero values $\Delta(\eta)$ of decreasing modes, they will grow beyond the bound for $\Delta(\eta_{in})$ at some finite conformal time $\eta_c$. Extrapolating further backwards the solution has to diverge if $\Delta(\eta)$ is predicted to be zero - otherwise there would be no such prediction. Such a singularity cannot be removed by a conformal transformation. It does not indicate a physical singularity either. It rather tells us that a finite value of $\Delta (\eta)$ is not allowed for a Universe lasting since ever. Not every inhomogeneous fluctuation in the neighborhood of the homogeneous solution can be extrapolated backwards without encountering a singularity at finite $\eta$. The backwards extrapolation has to respect the prediction for the decreasing modes. 

In summary, a beginning of the Universe in the infinite past \textit{predicts} that arbitrary field configurations at finite $\eta$ generically diverge when extrapolated backwards. This is the necessary consequence of the presence of decreasing modes. Backwards extrapolation to the infinite past is possible only for fields  within the predicted allowed range.
In contrast, if a non-zero amplitude of a decreasing mode would be observed, this would indicate that the corresponding cosmology cannot have lasted forever.

\bigskip

\textbf{Discussion.} \enspace
We have shown that 
the homogeneous isotropic cosmological solution for 
standard models of inflationary cosmology can be extrapolated backwards to the infinite past in physical time, as measured by the number of oscillations of photons. This extends to our observed inhomogeneous Universe for two alternative settings. Either the beginning of the Universe is inhomogeneous, with a homogeneous expectation value of the metric tending to zero in the infinite past. Only once the homogeneous expectation value increases to values larger than the inhomogeneities, the Universe becomes more and more homogeneous and the linear approximation for relative inhomogeneities becomes valid. Or the propagators for observable relative fluctuations remain finite.
No physical big bang singularity is present in either case. The often discussed singularity is then only apparent, being related to a singular, and therefore not very appropriate, choice of coordinates in field space. Field relativity permits us to use better adapted choices for the metric field. In particular, in the primordial flat frame the averaged geometry becomes flat Minkowski space in the infinite past. The absence of singularities for the homogeneous solution becomes very apparent.

The lightlike vacuum in the beginning of the Universe can be associated to quantum scale symmetry \cite{Wetterich2019}. Unbroken scale symmetry implies massless particles, as encountered in the lightlike vacuum. Quantum scale symmetry arises from an ultraviolet fixed point in the flow of couplings, functions or functionals in quantum gravity coupled to particle physics. For interesting ``crossover cosmologies" \cite{Wetterich2014a,Rubio2017} the Universe starts from an ultraviolet fixed point in the infinite past, and makes a transition or crossover to a different infrared fixed point that is approached in the infinite future.

We emphasize that a beginning as a lightlike vacuum is possible for inflationary cosmologies, but not mandatory. Other possible histories of the Universe, as a crossing of the apparent big bang singularity in a bouncing Universe \cite{Bars2013, Kamenshchik2016}, or emergence of our Universe from a finite region of a multiverse \cite{Linde1983, Shafi1983}, can be imagined. In this case the lightlike vacuum would not last forever towards the infinite past. It would rather be reached at some particular time characterizing the bounce or the onset of inflation for a region. Nevertheless, no necessity for such an extension is visible at present. For a long epoch in physical time the physical properties of the Universe can be characterized by great emptiness, independently of the detailed beginning and the issue of singularities.

\nocite{*}
\bibliography{GE_refs}

\end{document}